# Orientation-dependent stability and quantum-confinement effects of silicon carbide nanowires

Zhenhai Wang[1], Mingwen Zhao[1, 2, 3], Tao He[1], Hongyu Zhang[1], Xuejuan Zhang[1], Zexiao Xi[1], Shishen Yan[1, 2], Xiangdong Liu[1, 2], and Yueyuan Xia[1, 2]

[1] School of Physics, Shandong University, Jinan 250100, Shandong, China

[2] State Key Laboratory of Crystal Materials, Shandong University, Jinan, Shandong, 250100, China

E-mail: zmw@sdu.edu.cn

**Abstract**

The energetic stability and electronic properties of hydrogenated silicon carbide nanowires (SiCNWs) with zinc blende (3C) and wurtzite (2H) structures are investigated using first-principles calculations within density functional theory and generalized gradient approximation. The [111]-orientated 3C-SiCNWs are energetically more stable than other kinds of NWs with similar size. All the NWs have direct band gaps except the 3C-SiCNWs orientating along [112] direction. The band gaps of these NWs decrease with the increase of wire size, due to the quantum-confinement effects. The direct-band-gap features can be kept for the 3C-SiCNWs orientating along [111] direction with diameters up to 2.8 nm. The superior stability and electronic structures of the [111]-orientated 3C-SiCNWs are in good agreement with the experimental results.

---

[3] Author to whom any correspondence should be addressed.

# 1. Introduction

One-dimensional nanostructures with electrical carriers confined in the other two directions, have been drawing considerable interests because of its unique physical properties and potential applications in the construction of electronic, optical, electrochemical, and electromechanical devices. Silicon carbide (SiC) is particularly attractive due to its excellent mechanical properties, high chemical resistivity, and its widespread applications in high-temperature, high-voltage electronics and short-wavelength optics. Recently, one-dimensional SiC nanostructures with varied morphologies, such as nanowires [1-3], nanorods [4], nanobelts [5], nanotubes [6] and nanocables [7], have attracted extensive attention. Among these polytypes, SiC nanowires (SiCNWs) are of particular interest because of their distinctive physical and chemical properties and the superior quantum confinement in these quasi-one-dimensional nanowires [8], which make them excellent candidates for designing and fabricating nanodevices. Various approaches, such as carbon template [1, 9], arc discharge [10], and thermal evaporation [11] are available for producing SiCNWs. The as-synthesized SiCNWs always have a single-crystalline core of 3C form orientating along [111] direction. The quantum confinement of SiC nanostructures has been evidenced by the photoluminescence and photoluminescence excitation spectral examinations of the 3C-SiC nanocrystallites [12]. The emission band maximum ranging from 440 nm to 560 nm was attributed to the band-to-band recombination of photoexcited carriers in the quantum confined 3C-SiC nanocrystallites with the size of 1-6 nm.

Meanwhile, theoretical insights into the unique properties of SiC nanowires are crucial for the utilization of these nanostructures. The nanomechanical response properties of [111]-orientated 3C-SiCNWs under different types of loading strain have been investigated using molecular dynamics simulation with Tersoff bond-order interatomic potential [13]. The electronic structure and uniaxial-stress effects of small-size [111]-orientated 3C-SiCNWs with the diameter less than 1.23 nm were also studied by performing first-principles calculations within density-functional theory (DFT) [14]. Using an atomistic band-order potential in conjunction with DFT calculations, the energetic stability and electronic structures of wurtzite (2H) SiCNWs and SiC nanotubes with diameters ranging from 0.9 nm to 5.8 nm were predicted in our previous work [15]. However, to the best of our knowledge, the stability and electronic structures of SiCNWs orientating along other high-symmetric directions of 3C-SiC, such as [110] and [112], have not been reported. More importantly, previous works have shown the direct-band-gap features of [111]-orientated silicon NWs arising from quantum-confinement will be lost when the wire diameter is larger than 2.0 nm [16]. Whether the direct-band-gap features of [111]-orientated SiCNWs can be kept for large-size NWs or not remains unclear. In view of the potential applications of SiCNWs in nanoscience and nanotechnology, the study of the above issues is timely desirable and becomes the goals of the present work.

We performed first-principles calculations within DFT and generalized gradient approximation (GGA) to study the energetics and electronic structures of the SiCNWs orientating along [110], [111], [112] and [0001] directions with 3C and 2H forms. The diameters of the SiCNWs considered in this work are up to 2.8 nm. The size- and orientation-dependent quantum-confinement of SiCNWs revealed in this work is expected to promote the utilization of these nanostructures in building nanoscaled optoelectronic devices which is inaccessible for bulk SiC materials with indirect band gaps.

**2. Theoretical methods and computational details**

All of the first-principles calculations were performed using DFT and GGA exchange-correlation functional in the form of Perdew, Burke and Ernzerhof (PBE) [17] which is implemented in the SIESTA code [18-20]. A flexible linear combination of numerical atomic orbital basis sets of double-$\zeta$ quality with inclusion of polarization functions (DZP) was adopted for the description of valence electrons, and nonlocal pseudo-potentials were employed for the atomic cores. The basis functions were strictly localized within radii that corresponded to confinement energy of 0.01 Ry, with the exception of the polarization functions where a fixed radius of 6.0 Bohr was specified. An auxiliary basis set with a real-space grid was adopted to expand the electron density for numerical integration. A kinetic energy cutoff of 200 Ry was employed to control the fineness of this mesh. One-dimensional (1D) periodic boundary condition was applied along the axial direction and a sufficient vacuum space (up to 20 Å) was specified along the radial direction to avoid mirror interactions. The Brillouin zone sampling was carried out by using k-point grids of 1×1×8 according to the Monkhorst-Pack scheme [21]. All of the atomic positions along with the lattice vectors were optimized by using a conjugate gradient (CG) algorithm, until each component of the stress tensors was reduced below 0.02 GPa and the maximum atomic forces was less than 0.01 eV/Å. This scheme reproduces well the equilibrium structural parameters of bulk SiC crystals with the maximum deviation less than 1%. For example, the optimized lattice constant (a) for 3C-SiC crystal obtained from the present calculations are 4.37 Å, agree well with the experimental data 4.35 Å. For 6H-SiC crystal, the optimized lattice constants are a = 3.09 Å and c = 15.23 Å, close to the experimental results a = 3.08 Å and c = 15.12 Å. Similar scheme has also been employed in our previous work [15]

## 3. Results and discussion

The single-crystalline SiCNWs considered in this work are modeled by high symmetric prisms with axes orientating along [110], [111], and [112] directions of 3C-SiC crystal and [0001] direction of 2H-SiC crystal enclosed by low-index facets, as shown in Figure 1. These SiCNWs have different shapes of cross sections, e.g. rhombi for [110] NWs, hexagons for [111] and [0001] NWs, and rectangles for [112] NWs. The size of the NWs can be represented by the number of SiC units in per cell of the NWs. The dangling bonds on the facets are passivated by hydrogen atoms to stabilize the NWs. The equilibrium configurations shown in Figure 1 indicate that the atom arrangement in these NWs keeps the features of the corresponding bulk crystals (3C- or 2H-SiC). Quite different from the NWs with clean facets [15], no surface reconstruction is found in these hydrogenated SiCNWs. Surface hydrogenation is quite necessary to stabilize the facets, especially for those with high density of dangling bonds, such as the $(1\bar{1}0)$ surface. The average Si-H and C-H bond lengths on the facets are 1.499 Å and 1.106 Å, close to the values of $CH_4$ and $SiH_4$ molecules, 1.483 Å and 1.091 Å, indicating that these H atoms are chemically bonded onto the facets. Because the densities of dangling bonds are different for facets of these NWs, the stoichiometries differ from one NW to another. To evaluate the energetic stability of the SiCNWs with different stoichiometries, we define the formation energy $E_{form}$ of a hydrogenated SiCNW by the expression:

$$E_{form} = E_{tot} - n_{SiC}\mu_{SiC} - n_H\mu_H$$

where $E_{tot}$ is the total energy of the hydrogenated NW, $n_{SiC}$ and $\mu_{SiC}$ are the number and chemical potential of SiC units, respectively, $n_H$ and $\mu_H$ are the number and chemical potential of H atoms. It's well known that the chemical potential of a species highly depends on the experimental conditions, such as temperature, pressure and solution environment. In this work, the value of $\mu_{SiC}$ was calculated from bulk SiC crystals at 0 K, whereas $\mu_H$ varies around the half of the total energy of a $H_2$ molecule (-16 ~ -8 eV). This allows us to simulate the energetic stability of SiCNWs growing under different conditions.

Figure 2 gives the formation energies of different types of SiCNWs with close sizes as a function of $\mu_H$. Different from clean SiCNWs which have positive formation energies [15], all of the NWs have negative values of $E_{form}$, suggesting that surface hydrogenation can effectively stabilize these NWs. The values of $E_{form}$ per SiC unit decrease linearly with the increase of $\mu_H$. Clearly, the energetic stability of these SiCNWs is orientation-dependent. The [111]-orientated NWs with hexagonal cross sections are energetically the most favorable, followed by the NWs orientating along [112] direction with rectangular cross sections. The [110]-orientated SiCNWs are energetically the most unfavorable. The $E_{form}$ differences between the NWs with close sizes decrease with the increase of wire sizes represented by the number of SiC units in per supercell. This is related to the H/SiC ratios of these NWs which decrease with increasing wire sizes. The superior stability of [111]-orientated SiCNWs over other kinds of SiCNWs is in good agreement with the experimental findings that the already-synthesized SiCNWs always have a 3C structure with axes orientating along [111] direction [1-3,22,23].

The electronic properties of SiCNWs also depend on the size and orientation of the NWs. In contrast to the indirect-band-gap features of bulk 3C-SiC crystal, the 3C-SiCNWs orientating along [110] and [111] directions and the [0001]-orientated 2H-SiCNWs have direct band gaps at Γ points. The [112]-orientated 3C-SiCNWs have indirect band gaps, as shown in Figure 3. The direct-band-gaps of [110]-orientated SiCNWs and the indirect-band-gaps of [112]-orientated SiCNWs are understandable in terms of the effective-mass-approximation. Bulk 3C-SiC crystal has an indirect band gap with the valence band maximum (VBM) and conduction band minimum (CBM) located at Γ and X points, respectively. Similar to the case of bulk silicon crystal, 3C-SiC has six equivalent conduction band minima (CBM) on ±x, ±y, ±z axes with the transverse mass (0.25) less than the longitudinal mass (0.68). When a [110] NW is constructed, the energy level of the CBM at ± (1, 0, 0)2π/a and ± (0, 1, 0)2π/a are upshifted due to the 1-D confinement and the bands at ± (0, 0, 1)2π/a are projected onto Γ, exhibiting both the large mass and the small mass in the confinement plane. The amount of upshift of the conduction band minima at ± (1, 0, 0)2π/a and ± (0, 1, 0)2π/a is greater than those projected onto the Γ-point, because the effective masses toward the confined directions of the CBM at ± (1, 0, 0)2π/a and ± (0, 1, 0)2π/a are small transverse masses. This gives rise to a direct band gap at Γ point. For the cases of [112]-orientated SiCNWs, no CBM can be folded into the Γ point, and the NWs have indirect band gaps, similar to the 3C-SiC bulk crystal. This is consistent with our DFT calculations. It is noteworthy that according to the effective masses approximation, [111]-orientated 3C-SiCNWs are expected to have indirect band gaps, similar to the cases of [112] SiCNWs. However, our DFT calculations clearly show that [111]-orientated SiCNWs considered in this work have direct band gaps at Γ point. The direct band gaps of [111]-orientated SiCNWs arise from the quantum-confinement-effects in these nanowires. The direct band gap features of [111]-orientated NWs have also been found for silicon nanowires[16,24,25]. Both experimental [24] and theoretical [16] works showed that silicon nanowires growing along [111] direction have direct band gaps when the wire diameter (d) is smaller than 2 nm, and those with d > 2.0 have indirect band gaps. We calculated the electronic bands of [111]-orientated 3C-SiCNWs with diameter as large as 2.8 nm, and didn't find the transformation from direct to indirect band gap. This implies that the direct band gap features of [111]-orientated SiCNWs can be kept in a wider range of size, compared to the silicon nanowires. Because of the computation limitation, the SiCNWs with diameter larger than 2.8 nm are not calculated in this work. The size of indirect-direct band gap transition remains unclear at present.

Due to the quantum-confinement effects, the band gaps of these SiCNWs decrease with the increase of wire size, as shown in Figure 4. The band gaps ($E_{gap}$) can be fitted using the expression:

$$E_{gap} = E_g^{bulk} + \beta \times (n_{SiC})^{-\alpha}$$

where $E_g^{bulk}$ is the band gap values of the corresponding bulk (3C or 2H) crystal obtained from the present calculations, $\alpha$ and $\beta$ are the fitting parameters, $n_{SiC}$ is the number of SiC units in per cell. Obviously, for a certain kind of SiCNWs, $(n_{SiC})^{1/2}$ is proportional to the radial size. The fitting data for these SiCNWs are: α = 0.58, β = 5.51 for the [110] NWs, α = 0.60, β = 12.7 for the [111] NWs, α = 0.63, β = 11.4 for the [112] NWs, and α = 0.73, β = 9.89 for the [0001] NWs. The value α = 1 is expected using an effective-mass particle-in-a-box approach [26]. This deviation is due to the quantum-confinement effects of these NWs. Similar results have also been found for the band gaps of [111]-orientated silicon NWs which have α = 0.85 [16]. Moreover, the band gap values of 3C-SiCNWs orientating along [111] direction (2.81 eV - 2.68 eV) are consistent with the emission band maximum ranging from 440 nm to 560 nm [12].

To reveal the origination of the bands near the Fermi level, we calculated the electron wave functions of the states corresponding to the VBM and CBM at Γ and X points of [111]-orientated 3C-SiCNWs. The isosurfaces of these wave functions are plotted in Figure 5. The wave function isosurfaces corresponding to VBM and CBM mainly distribute around the carbon and silicon atoms in the core region of the wire, respectively. This is also consistent with the electron density of states projected onto the atoms of the wire which shows that the electron states at VBM and CBM mainly originate from the C (2p) and Si (3p) states in the core region, respectively. Surface states are removed from the region near the Fermi level. This differs significantly from the bared SiCNWs where the electron bands near the Fermi level arise from the surface states [15, 27].

## 4. Conclusion

Our first-principles study of SiCNWs growing along different directions with zinc blende and wurtzite structures shows that the SiCNWs growing along [111] direction with zinc blended form are energetically most favorable, in good agreement with the experimental results. In contrast to the indirect band gaps of bulk crystals, all the NWs have direct band gaps at $\Gamma$ points, except the ones orientating along [112] directions which have indirect band gaps. The direct-band-gap features of [111]-orientated SiCNWs arising from the quantum-confinement effects can be kept for the NWs with diameters up to 2.8 nm. The band gap values decrease with the increase of wire diameters.

**Acknowledgement**

This work is supported by the National Natural Science Foundation of China under grant No. 10675075, the National Basic Research 973 Program of China (Grant No. 2005CB623602).

**Table 1.** The diameter (*d*), number of SiC units ($n_{SiC}$) and number of hydrogen atoms ($n_H$), and the band gap values ($E_g$) of the SiC nanowires orientating along different directions. The values of *d* are measured by the average double distance between the wire axis and the hydrogen atoms. The band gap values corrected by adding a size independent constant obtained from bulk (0.93eV for 3C-NWs and 0.9 eV for 2H-NWs) are listed in the brackets.

| Directions | | *d* (nm) | $n_{SiC}$/cell | $n_H$/cell | $E_g$ (eV) |
|---|---|---|---|---|---|
| [110] | NW1 | 0.88 | 8 | 12 | 3.03 (3.96) |
| | NW2 | 1.53 | 24 | 20 | 2.25 (3.18) |
| | NW3 | 2.16 | 48 | 28 | 1.95 (2.88) |
| | NW4 | 2.78 | 80 | 36 | 1.82 (2.75) |
| [111] | NW1 | 0.55 | 7 | 18 | 5.35 (6.28) |
| | NW2 | 0.88 | 19 | 30 | 3.57 (4.50) |
| | NW3 | 1.20 | 37 | 42 | 2.90 (3.83) |
| | NW4 | 1.53 | 61 | 54 | 2.44 (3.37) |
| | NW5 | 1.85 | 91 | 66 | 2.23 (3.16) |
| | NW6 | 2.18 | 127 | 78 | 2.06 (2.99) |
| | NW7 | 2.44 | 169 | 90 | 1.97 (2.91) |
| | NW8 | 2.77 | 217 | 102 | 1.88 (2.81) |
| [112] | NW1 | 0.70 | 8 | 16 | 4.46 (5.39) |
| | NW2 | 1.16 | 24 | 28 | 2.98 (3.91) |
| | NW3 | 1.63 | 48 | 40 | 2.38 (3.31) |
| | NW4 | 2.08 | 80 | 52 | 2.05 (2.98) |
| [0001] | NW1 | 0.55 | 6 | 12 | 5.43 (6.33) |
| | NW2 | 1.18 | 24 | 24 | 3.85 (4.75) |
| | NW3 | 1.75 | 54 | 36 | 3.29 (4.19) |
| | NW4 | 2.32 | 96 | 48 | 3.02 (3.92) |

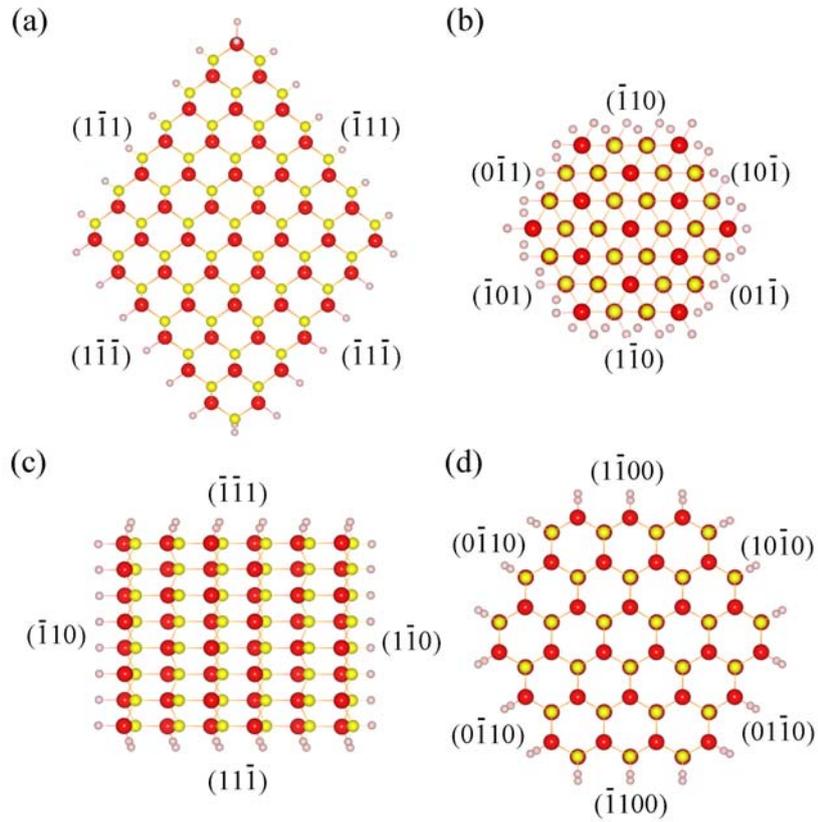

Figure 1. Top view of the relaxed configurations of SiC nanowires (NW3) enclosed by low-index facets along different directions. (a) [110], (b) [111] and (c) [112] are 3C polytypes, while (d) [0001] is 2H polytype. C, Si and H atoms are represented by yellow, red, and white spheres.

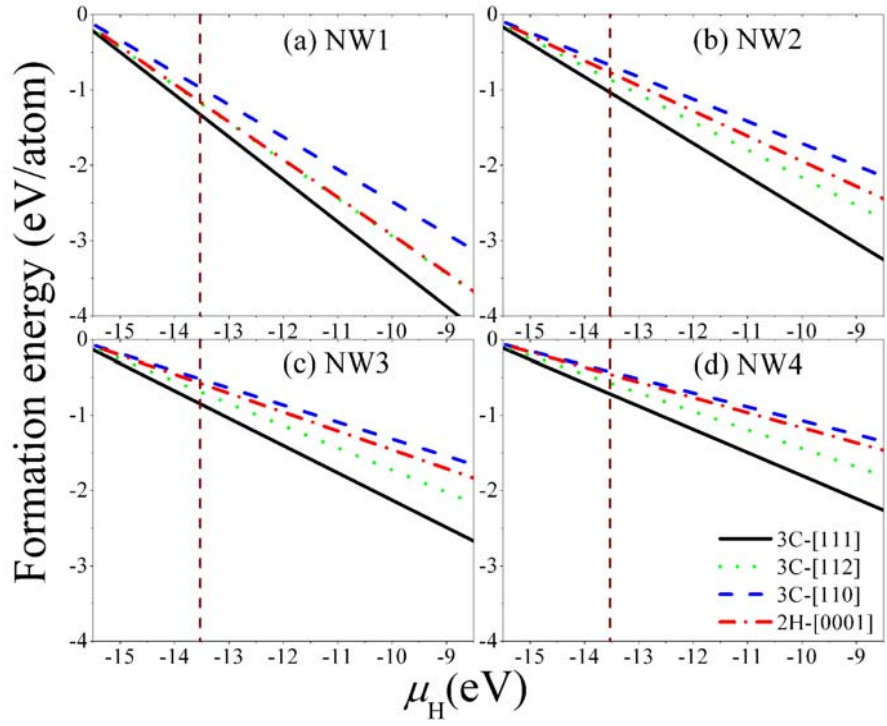

Figure 2. The formation energies of SiC nanowires orientating along different orientations as a function of H chemical potential ($\mu_H$). The diameters of these nanowires are listed in Table 1. The chemical potential of a hydrogen atom in vacuum calculated from the present calculations (-13.53 eV) is indicated by the vertical dashed lines of these figures.

Figure 2 Wang *et al.*,

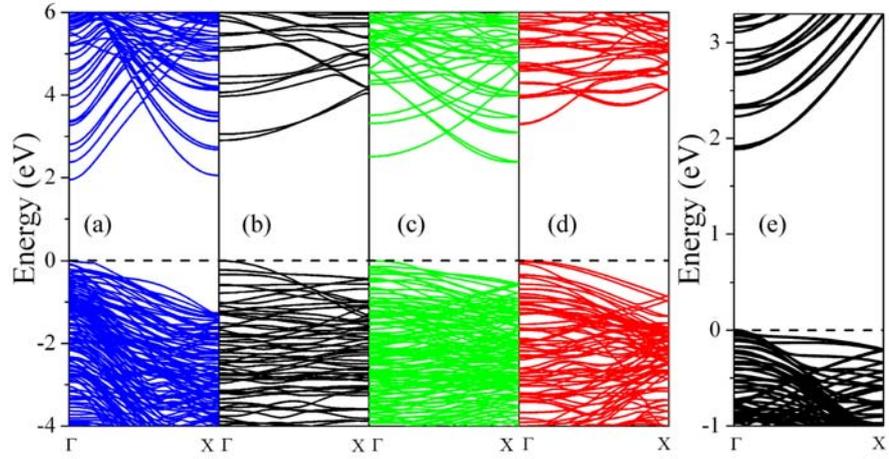

Figure 3. The electronic band structures of SiC nanowires (NW3) orientating along (a) [110], (b) [111], (c) [112], and (d) [0001] directions. The atomic structures of these nanowires are shown in Figure 1. (e) The electron bands of the largest 3C-SiCNW (NW8). The energy at Fermi level is set to zero.

Figure 3 Wang *et al.*,

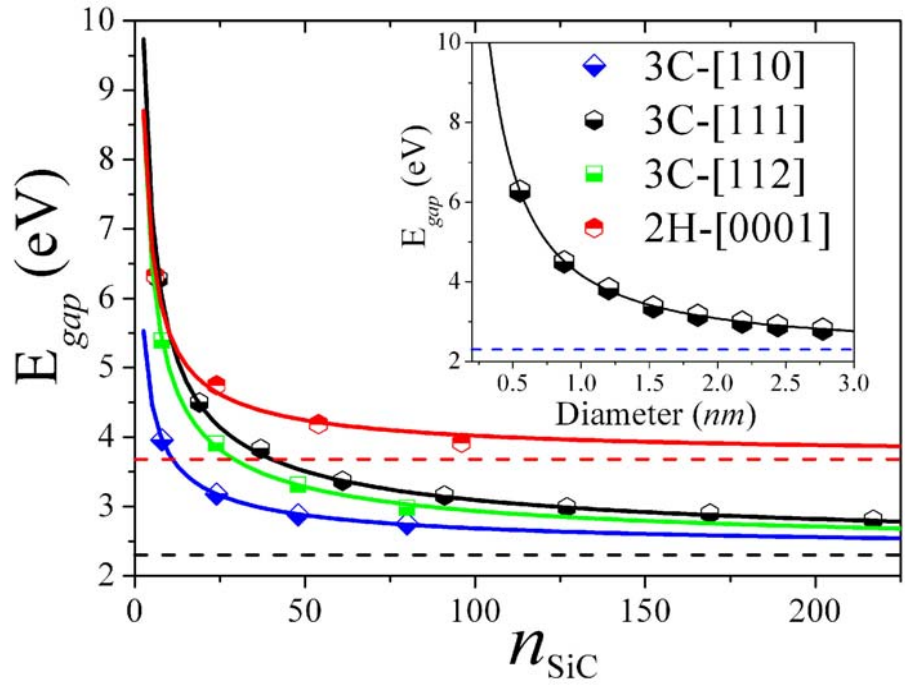

Figure 4. Band gap values of SiC nanowires orientating different directions as a function of nanowire size represented by the number of SiC units ($n_{SiC}$). The solid lines are the fitting data given by the expression: $E_{gap} = E_g^{bulk} + \beta \times (n_{SiC})^{-\alpha}$. The inset shows band gaps of 3C [111]-oriented NWs as a function of diameter (*d*): $E_{gap} = 2.3 + 1.87 \times d^{-1.27} (eV)$

Figure 4 Wang *et al.*,

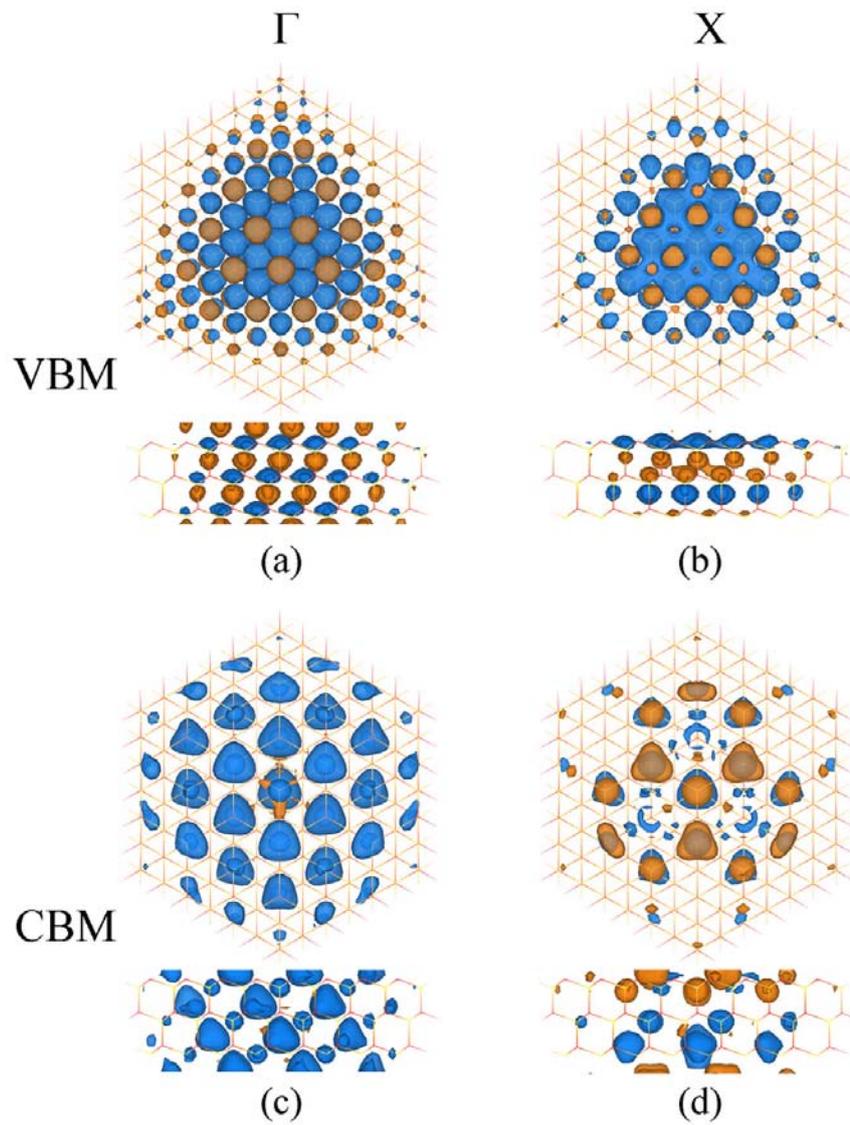

Figure 5. The isosurfaces of the Kohn-Sham states of (a), (b) the valence band maximum (VBM) and (c), (d) the conduction band minimum (CBM) at Γ and X points of one 3C-SiCNW orientating along [111]-direction. The isovalue is 0.03 Å$^{-3/2}$.

Figure 5 Wang *et al.*